# ABLA07 - towards a complete description of the decay channels of a nuclear system from spontaneous fission to multifragmentation

Aleksandra Kelić, M. Valentina Ricciardi and Karl-Heinz Schmidt
*GSI, Planckstrasse 1, D-64291 Darmstadt, Germany*

## Abstract

The physics and the technical algorithms of the statistical de-excitation code ABLA07 are documented. The new developments of ABLA07 have been guided by the empirical knowledge obtained in a recent experimental campaign on the nuclide distributions measured at GSI, Darmstadt. Besides distinct signatures of very asymmetric binary splits, lighter systems show clear features of multifragmentation, while heavy systems reveal the influence of dynamics and microscopic structure on the fission process. ABLA07 includes elaborate but efficient descriptions of all these processes, with one set of the model parameters fixed for all systems and all energies.

TABLE OF CONTENTS

# I. Introduction

Nuclear reactions represent an excellent tool to study static and dynamical properties of nuclear matter. For example, while fission at low excitation energies can be used to extract information on the heights and curvatures of the fission barriers [1] or on pairing and shell correlations at large deformations [2,3], fission at high excitation energies (above ~100 MeV) can give insight into dissipative properties of nuclear matter, see e.g. [4]. Other examples are spallation reactions and heavy-ion reactions at and above the Fermi energy, which are best suited for studying thermal instabilities and the liquid-gas phase transition in nuclear matter [5,6,7,8,9,10]. Unfortunately, most of the needed information cannot be directly obtained from the experimental observables. Usually, one needs to use some reaction model, and only by comparing the predictions of a considered model with measured observables one can gain more insight into the physical processes. For this purpose, of course, reaction models as realistic as possible are mandatory.

In recent years, reaction models became important not only for the basic research but also for different applications. Fusion, fission, fragmentation or spallation reactions are used in order to produce beams of secondary, radioactive ions. Spallation reactions are used as a neutron source [11,12] and considered for different applications such as for example nuclear-waste management [13,14]. To these purposes, many facilities are being built or being planned all around the world. For the proper functioning of these facilities, cross sections of different particles produced in considered nuclear reactions have to be known. Due to the variety of the involved systems, i.e. different target/projectile/beam-energy combinations, not all production cross sections can be measured, and one has to rely on model calculations. Therefore, reliable and fast reaction models are also mandatory for technological applications.

Usually, nuclear-reaction models consist of two stages: In the first stage, it is assumed that due to the interaction between a target and a projectile nucleus, an excited thermally equilibrated nuclear system is formed. After the thermalised system is formed, in the second stage its decay is described in the frame of the statistical model [15]. Usually, two realisations of the statistical model are employed: The Weisskopf-Ewing approach [16] and the Hauser-Feshbach approach [17]. While in the former a direct consideration of angular momentum and parity is neglected, in the latter approach they are explicitly taken into account. Many different deexcitation codes based on these two approaches have been developed. They mostly differ according to different descriptions of the physics concepts involved, e.g. level density, nuclear potential, nuclear viscosity or number of considered decay channels. In order to be used for the description of the deexcitation process of residues formed in different types of interactions (i.e. different target/projectile/energy combinations) a deexcitation code has to be adapted to some specific needs:

- A consistent treatment of level densities as a function of excitation energy and nuclear shape is mandatory. The treatments of shell effects [18] and collective excitations [19] are particularly important.

- The dynamics of the fission process and the onset of thermal instabilities at the highest temperatures have to be considered.

- Modelling of fission requires considering a large variety of fissioning nuclei in a wide range of excitation energies. Available empirical formulations of nuclide distributions in fission of specific nuclei should be replaced by a model, which is based on more fundamental properties, like the potential-energy landscape around saddle and scission.

- For application purposes, inclusion in complex transport codes demands for short computing times.

In the following, we will describe the deexcitation code ABLA07, which complies with the above-mentioned requirements.

## II. Description of the model

ABLA07 is a dynamical code that describes the de-excitation of the thermalised system by simultaneous break-up, particle emission and fission. Simultaneous break-up is considered as the cracking of the hot nucleus into several fragments due to thermal instabilities. The description of particle evaporation is based on the Weißkopf-Ewing formalism [20], while the fission decay width is calculated taking into account dynamical effects [21]. The basic ingredients of the model are[1]:

1. Emission of neutrons, light charged particles ($Z=1, 2$), intermediate-mass fragments IMF ($Z>2$) and gamma rays is considered.

2. In calculating the particle decay widths the following effects are considered:

   - Energy dependent inverse cross sections based on nuclear potential using the ingoing-wave boundary condition model [22].

   - Barriers for charged particles are calculated using the Bass potential [23].

   - Thermal expansion of the source [24] is taken into account.

   - Change of angular momentum due to particle emission is considered.

3. The fission decay width is described by including:

   - An analytical time-dependent approach [25,26] to the solution of the Fokker-Planck equation,

   - The influence of the initial deformation on the fission decay width,

   - The double-humped structure in the fission barriers of actinides,

   - Symmetry classes in low-energy fission.

4. Particle emission on different stages, i.e. between ground state and saddle point, between the saddle and scission point, and from two separate fission fragments, of the fission process is calculated separately.

5. Kinetic-energy spectra of the emitted particles are directly calculated from the inverse cross sections.

6. A stage of simultaneous break-up [9] in the decay of hot excited systems is explicitly treated.

In the following, these different steps will be discussed in more details.

### II.1. Particle emission

Following the Weißkopf-Ewing formalism [20,27], the decay width of a specific initial nucleus, characterised by its excitation energy $E_i$ into a daughter nucleus with excitation energy $E_f$ by emission of particle $\nu$ with kinetic energy $\varepsilon_\nu$ is given as:

$$\Gamma_\nu(E_i) = \frac{2 \cdot s_\nu + 1}{2 \cdot \pi \cdot \rho_i(E_i)} \cdot \frac{2 \cdot m_\nu}{\pi \cdot \hbar^2} \cdot \int_0^{E_i - S_\nu - B_\nu} \sigma_c(\varepsilon_\nu) \cdot \rho_f(E_f) \cdot (\varepsilon_\nu - B_\nu) dE_f \quad (1)$$

---

[1] Comparison with the previous version of the model ABLA is given in Annex I.

In the above equation, $s_\nu$ is the spin of the emitted particle, $\rho_i$ and $\rho_f$ are the level densities in the initial and the daughter nucleus, respectively, $\sigma_c$ is the cross section for the inverse process, $B_\nu$ is the Coulomb barrier for charged-particle emission and $m_\nu$ the mass of the emitted particle.

In order to calculate the probability of a certain decay channel, i.e. $P_\nu = \Gamma_\nu / \sum \Gamma_i$, one needs, therefore, several important parameters: the level density, the Coulomb barrier and the inverse cross section. Below, we discuss them in more details.

### II.1.1. Level density

The total level density used in Eq. (1) is calculated as the product of the intrinsic level density $\rho_{in}(E)$ and the vibrational and rotational enhancement factors, $K_{vib}(E_{corr})$ and $K_{rot}(E_{corr})$, respectively [28]:

$$\rho(E) = \rho_{in}(E) \cdot K_{vib}(E) \cdot K_{rot}(E). \tag{2}$$

The intrinsic density of excited states, $\rho_{in}$, is calculated with the well-known Fermi-gas formula:

$$\rho_{in}(E) = \frac{\sqrt{\pi}}{12} \frac{\exp(S)}{\tilde{a}^{1/4} E_{eff}^{5/4}}, \tag{3}$$

with the exponent $S$:

$$S = 2 \cdot \sqrt{\tilde{a} \cdot E_{corr}} = 2 \cdot \sqrt{\tilde{a} \cdot \left(E_{eff} + \delta U \cdot k(E_{eff}) + \delta P \cdot h(E_{eff})\right)}, \tag{4}$$

and the asymptotic level-density parameter $\tilde{a}$ as given in Ref. [18]:

$$\tilde{a} = 0.073 \cdot A + 0.095 \cdot B_S \cdot A^{2/3}, \tag{5}$$

where $A$ is the mass of the nucleus, and $B_s$ is the ratio between the surface of the deformed nucleus and a spherical nucleus. $\delta U$ is the shell-correction energy, which is for the ground state calculated according to Ref. [29]. At the fission saddle point, the shell-correction energy is assumed to be negligible [30,31]. The function $k(E_{eff})$ describes the damping of the shell effect with excitation energy, and is calculated according to Ref. [18] as $k(E_{eff}) = 1 - \exp(-\gamma E_{eff})$, with the parameter $\gamma$ determined by $\gamma = \tilde{a} / (0.4 \cdot A^{4/3})$ [32].

The parameter $\delta P$ of equation (4), which is identical to the pairing condensation energy in odd-odd nuclei, is calculated as:

$$\delta P = -\frac{1}{4} \cdot \Delta^2 \cdot g + 2 \cdot \Delta, \tag{6}$$

with an average pairing gap $\Delta = 12 / \sqrt{A}$, and the single-particle level density at the Fermi energy $g = 6 \cdot \tilde{a} / \pi^2$. The function $h(E_{eff})$ parameterises the superfluid phase transition [33] at the critical energy $E_{crit} = 10$ MeV [34]:

$$h(E_{eff}) = \begin{cases} 1 - \left(1 - \dfrac{E_{eff}}{E_{crit}}\right)^2, & E_{eff} < E_{crit} \\ 1, & E_{eff} > E_{crit} \end{cases}. \tag{7}$$

The effective energy $E_{eff}$ is shifted with respect to the excitation energy $E$ to accommodate for the different energies of even-even, odd-mass, and odd-odd nuclei:

$$E_{eff} = E \qquad \text{odd } Z - \text{odd } N$$

$$E_{eff} = E - \Delta \qquad \text{odd } A$$

$$E_{eff} = E - 2\Delta \qquad \text{even } Z - \text{even } N.$$

In order to calculate the intrinsic level density at very low excitation energies, we switch from the Fermi-gas level density to the constant-temperature level density [35]. The calculation is based on the work performed in Ref. [36], where the values of the parameters of the constant-temperature level density approach were obtained from the simultaneous analysis of the neutron resonances and the low-lying levels in the framework of the Gilbert-Cameron approach [35].

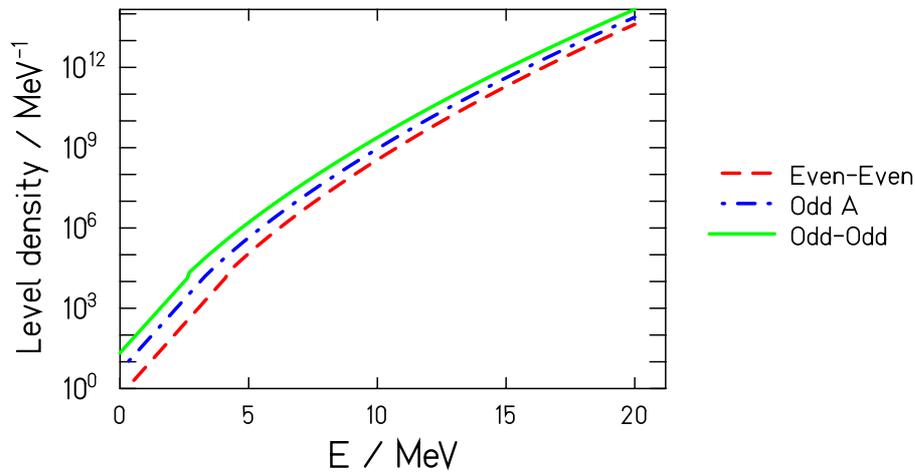

Figure 1: Intrinsic level density $\rho_{in}(E)$ for three nuclei – $^{242}$Cf, $^{241}$Bk and $^{240}$Bk – calculated in ABLA07 using combined Fermi-gas – constant-temperature level density approach.

As it was shown in Ref. [19], collective excitations can contribute considerably to the total nuclear level density. In deformed nuclei, the most important contribution to the collective enhancement of the level density originates from rotational bands, while in spherical nuclei the collective enhancement is caused by vibrational excitations.

In ABLA07, the contribution of collective excitations to the level density is described in the following way (for more details, see Ref. [19]): For nuclei with a quadrupole deformation $|\beta_2| > 0.15$, the rotational enhancement factor $K_{rot}(E_{corr})$ is calculated in terms of the spin-cutoff parameter $\sigma_\perp$:

$$K_{rot}(E_{corr}) = \begin{cases} (\sigma_\perp^2 - 1) \cdot f(E_{corr}) + 1, & \sigma_\perp^2 > 1 \\ 1, & \sigma_\perp^2 < 1 \end{cases} \qquad (8)$$

$$\sigma_\perp^2 = \frac{\Im_\perp \cdot T}{\hbar^2}, \quad f(E_{corr}) = \left(1 + \exp\left(\frac{E_{corr} - E_c}{d_c}\right)\right)^{-1}, \qquad (8a)$$

where $E_{corr}$ is defined in Eq. (4), $\Im_\perp = \frac{2}{5} m_0 \cdot A \cdot R^2 (1 + \beta_2/3)$ is the rigid-body moment of inertia perpendicular to the symmetry axis, and $m_0$ is the mass unit. The ground-state quadrupole deformation $\beta_2$ is taken from the finite-range liquid-drop model including microscopic corrections [29], while the saddle-point deformation is taken from the liquid-drop model as given in Ref. [37]. The damping of the collective modes with increasing excitation energy is described by a Fermi function $f(E)$ with parameters $E_c = 40$ MeV and $d_c = 10$ MeV. The vibrational enhancement for spherical nuclei is generally smaller than the rotational enhancement for deformed nuclei. For nuclei with a quadrupole deformation $|\beta_2| < 0.15$, the vibrational enhancement factor is calculated by using the same formula as for the rotational enhancement (Eq. (8)), but with the spin-cutoff parameter which is, in order to simulate the vibrational motion, calculated assuming irrotational flow: $\sigma'^2_\perp = 70 \cdot \beta^2_{eff} \cdot \sigma^2_\perp$, where $\sigma_\perp$ is given by Eq. (8a), and $\beta_{eff}$ is a dynamical deformation parameter: $\beta_{eff} = 0.022 + 0.003 \cdot \Delta N + 0.005 \cdot \Delta Z$; $\Delta N$ and $\Delta Z$ are the absolute values of the number of neutrons and protons, respectively, above or below the nearest shell closure.

### II.1.2. Influence of angular momentum

In the standard Weisskopf-Ewing approach, the change of angular momentum in the evaporation process due to particle emission is not treated. To overcome this limitation, we have developed a dedicated formalism, which calculates the distribution of orbital angular momentum in the emission of nucleons and fragments from excited nuclei with finite angular momentum.

The emitting (mother) nucleus with mass number $A_m$ has the angular momentum $l_m$ and the excitation energy $E_m^*$. After the emission of a fragment with mass number $A_f$, separation energy $S_f$, kinetic energy $K_f$, excitation energy $E_f^*$ and angular momentum $l_f$, the daughter nucleus with mass number $A_d$, angular momentum $l_d$ and excitation energy $E_d^*$ is formed.

In the classical approximation, the probability for the emission of the fragment with a given orbital angular momentum is determined by the phase space available for the daughter nucleus and the fragment after the fragment emission. Due to energy conservation, we have the following relation:

$$E_d^* + E_f^* = E_m^* + E_m^{rot} - E_d^{rot} - E_f^{rot} - E_{orb} - S_f. \qquad (9)$$

Here, only $E_d^{rot}$, $E_f^{rot}$ and $E_{orb}$ depend on the orbital angular momentum of the fragment. The optimum combination of final intrinsic and orbital angular momentum is defined by the collinear combination of the angular momenta:

$$|l_{orb}| = |l_m| - |(l_d + l_f)|,$$

which leads to the final configuration with the largest number of states in the final nucleus.

The number of final states is approximately given by:

$$\rho \propto (l_d + l_f) \cdot \exp\left(2 \cdot \sqrt{a\left(E_m^* - S_f - \frac{l_{orb}^2}{2\Theta_{orb}} - \frac{(l_d + l_f)^2}{2\Theta_m}\right)}\right). \qquad (10)$$

Renaming $(l_d + l_f) = l_{final}$, we get

$$\ln(\rho) \propto \ln(l_{final}) + 2 \cdot \sqrt{a\left(E_m^* - S_f - \frac{l_{orb}^2}{2\Theta_{orb}} - \left(\frac{l_m^2 - 2l_m l_{orb} + l_{orb}^2}{2\Theta_m}\right)\right)} \qquad (11)$$

- *The most probable value of the orbital momentum:*

To obtain the most probable value of $l_{orb}$ we search for the maximum of the function in Eq. (11). The full distribution given by Eq. (11) is well approximated by a Gaussian function, whose width is related to the second derivative of the distribution in (11).

For relativistic nucleus-nucleus and nucleon-nucleus collisions, mostly considered here, the value of $l_{final}$ is expected to be very close to $l_m$. This is why we expand the above function around $l_m$:

$$\ln(\rho_{approx}) \propto \frac{l_{orb}}{l_m} + 2 \cdot \sqrt{S_4} \cdot \left(1 - \frac{a}{2 \cdot S_4}\left(\frac{l_{orb}^2}{2\Theta_{orb}} - \frac{l_m l_{orb}}{\Theta_m} + \frac{l_{orb}^2}{2\Theta_m}\right)\right),$$

with $S_4 = a\left(E_m^* - S_f - \frac{l_m^2}{2\Theta_m}\right)$, which does not depend on $l_{orb}$.

Since $\Theta_m \gg \Theta_{orb}$, we can approximate:

$$\ln(\rho_{approx}) \propto \frac{l_{orb}}{l_m} + 2 \cdot \sqrt{S_4} \cdot \left(1 - \frac{a}{2 \cdot S_4}\left(\frac{l_{orb}^2}{2\Theta_{orb}} - \frac{l_m l_{orb}}{\Theta_m}\right)\right).$$

The maximum of this function is given by requiring the first derivative to be zero:

$$\frac{1}{l_m} - \frac{2\sqrt{S_4}}{2}\frac{a}{2}\frac{2 l_{orb}}{S_4 \Theta_{orb}} + \frac{2\sqrt{S_4}}{2}\frac{a l_m}{S_4 \Theta_m} = 0$$

or

$$\frac{1}{l_m} - \frac{a}{\sqrt{S_4}}\frac{l_{orb}}{\Theta_{orb}} + \frac{a}{\sqrt{S_4}}\frac{l_m}{\Theta_m} = \frac{1}{l_m} - \frac{a}{\sqrt{S_4}}\left(\frac{l_{orb}}{\Theta_{orb}} - \frac{l_m}{\Theta_m}\right) = 0$$

From this we determine the optimum value of $l_{orb}$:

$$l_{orb} = \Theta_{orb} \cdot \left(\frac{l_m}{\Theta_m} + \frac{\sqrt{S_4}}{a} \cdot \frac{1}{l_m}\right) \qquad (12)$$

The values of $\Theta_m$ and $\Theta_{orb}$ can be estimated as follows:

$$\Theta_m = \frac{2}{5}m_0 A_m r^2 = \frac{2}{5}m_0 r_0^2 A_m^{5/3}$$

$$\Theta_{orb} = m_0\left(A_1 d_1^2 + A_2 d_2^2\right) = m_0\left(A_1\left(\frac{A_2}{A_m}d\right)^2 + A_2\left(\frac{A_1}{A_m}d\right)^2\right) = m_0\left(\frac{d}{A_m}\right)^2\left(A_1^2 A_2 + A_1 A_2^2\right)$$

with $d = r_0\left(A_1^{1/3} + A_2^{1/3}\right)$ in the touching-sphere configuration.

- *The width of the orbital-momentum distribution:*

To estimate the width of the orbital-momentum distribution in one evaporation step, we first write down the second derivative of $\ln(\rho)$:

$$\frac{\partial^2 (\ln \rho)}{\partial l_{orb}^2} = \frac{a}{\sqrt{S_4} \cdot \Theta_{orb}}.$$

From this we conclude that the orbital-momentum distribution can be approximated by Gaussian with the standard deviation:

$$\sigma = \sqrt{\frac{\sqrt{S_4} \Theta_{orb}}{a}}. \qquad (13)$$

In an evaporation code, the emission of a particle induces a change in angular momentum of the mother nucleus. This change is calculated by randomly picking the angular-momentum value from a Gaussian distribution with the mean value $l_{orb}$ (Eq. (12)) and the standard deviation $\sigma$ (Eq.(13)).

The most severe approximation in the above consideration is the restriction to collinear angular momenta in the evaporation process. This approximation is most crucial for the estimation of the widths of the orbital angular-momentum distribution, which may be underestimated. However, the most important value for the evaporation process is the most probable orbital angular momentum, which is only little affected, due to the dominant influence of the strong energy dependence of the level density.

### II.1.3. Inverse cross sections

In calculating the inverse cross section for the emission of particles one has to consider several effects: The existence of the Coulomb barrier for charged particles (especially at low energy), the tunnelling through it (especially for light particles), and the energy-dependent quantum-mechanical cross section.

At energies well above the Coulomb barrier the shape of the barrier does not play any role. $\sigma_c$ is then calculated without taking into account the tunnelling:

$$\sigma_c(\varepsilon_v) = \pi \cdot R^2 \cdot \left(1 - \frac{B_v}{\varepsilon_v}\right), \quad R = R_{geom} + R_\lambda \qquad (14)$$

$$R_{geom} = 1.16 \cdot \left(A_1^{1/3} + A_2^{1/3}\right), \quad \text{and} \quad R_\lambda = \sqrt{\frac{\hbar^2}{2 \cdot \mu \cdot E_{cm}}},$$

where $\mu$ is the relative mass (= $M_1 \cdot M_2/(M_1+M_2)$) and $E_{cm} = \varepsilon_v \cdot (A_1-A_2)/A_1$. $R_\lambda$ is obtained for the square-well potential and is responsible for the dependence of the capture cross section on the particle energy[2].

### II.1.4. Barriers for charged-particle emission

To calculate the Coulomb barrier, we use the nuclear potential for $l = 0$ ($V(r) = V_N(r) + V_C(r)$) and then numerically search for the position of the maximum that corresponds to the barrier.

---

[2] For low particle kinetic energy the wavelength associated to the particle becomes comparable to the nuclear dimensions, which results in the dependence of the cross section on particle energy.

- *The empirical nuclear potential of R. Bass* [38,39]:

$$-V_N(s) = \frac{C_1 \cdot C_2}{C_1 + C_2} \cdot \frac{1}{A \cdot \exp\left(\frac{s}{d_1}\right) + B \cdot \exp\left(\frac{s}{d_2}\right)}, \quad (15)$$

with the parameters:

$A = 0.333$ MeV$^{-1}$ fm, $\quad\quad\quad B = 0.007$ MeV$^{-1}$ fm,

$d_1 = 3.5$ fm, $\quad\quad\quad d_2 = 0.65$ fm.

$C_1$ and $C_2$ are the half-density radii of the daughter nucleus and emitted particle, respectively, calculated as:

$$C_i = R_i \cdot \left(1 - \frac{(0.9984\,\text{fm})^2}{R_i^2}\right), \quad R_1 = \left(1.28 \cdot A_f^{1/3} - 0.76 + \frac{0.8}{A_f^{1/3}}\right)\text{fm},$$

$$R_2 = \left(1.28 \cdot A_2^{1/3} - 0.76 + \frac{0.8}{A_2^{1/3}} + d\right)\text{fm}, \quad d = \begin{cases} 3\,\text{fm}, & 1\text{H} \\ 0\,\text{fm}, & 2\text{H} \\ 0\,\text{fm}, & 3\text{H} \\ 0\,\text{fm}, & 3\text{He} \\ 1\,\text{fm}, & 4\text{He} \end{cases}.$$

The variable $s = r - C_1 - C_2$ gives the distance between the two surfaces based on half-density radii.

- *Coulomb potential* [39]:

$$V_C = \begin{cases} 1.44 \cdot \dfrac{Z_1 \cdot Z_2}{r}, & r > R_C \\ 1.44 \cdot \dfrac{Z_1 \cdot Z_2}{2 \cdot R_C} \cdot \left(3 - \dfrac{r^2}{R_C^2}\right) & r \leq R_C \end{cases}, \text{ with } R_C = 1.3 \cdot \left(A_1^{1/3} + A_2^{1/3}\right) \text{ fm.} \quad (16)$$

Please note that inclusion of Eqs. (14-16) into the expression (1) for the particle decay width implies the use of numerical tools for solving the integral in Eq. (1) which can considerably increase the computational time. In order to overcome this problem, we have approximated the integrand in the Eq. (1) with a function, which allows us to analytically solve the integral. Details are given in Annex B.

### II.1.5. Tunnelling through the barrier

At energies below and just above the Coulomb barrier, the tunnelling of charged particles through the barrier plays an important role, and, consequently, the expression for the inverse cross section given by Eq. (14) is not any more applicable. In order to incorporate the effect of the tunnelling through the potential barrier, we follow the work done by Avishai in Ref. [40]. He considered two different energy ranges for calculating the inverse cross sections:

- *Energy below the Coulomb barrier:*

Avishai [40] showed that the nucleus-nucleus fusion cross section at sub-barrier energies can be predicted by the simple theory of Wong based on the barrier-penetration technique [41], where it is assumed that the reaction occurs whenever the two nuclei have penetrated through the potential barrier. For every angular momentum, i.e. every impact parameter, the penetration probability can be calculated by the Hill-Wheeler formula [42], after approximating the shape of the barrier by an inverted (half) parabola plus a Coulomb slope ($V \sim 1/r$).

If $R_l$ is the position of the top of the barrier, $E_l$ is the value of the effective interaction at its maximum and $\hbar \omega_L$ the curvature, the transmission coefficient for angular momentum $l$ can be calculated as:

$$P(l, \varepsilon_v) = \left\{ 1 + \exp\left[ \pi \cdot \frac{E_l - \varepsilon_v}{\hbar \omega_l} + C(\varepsilon_v) \right] \right\}^{-1}. \tag{17}$$

$C(E)$ express the penetration through the Coulomb part. Once the penetration coefficients of Eq. (17) are summed over all the possible angular momenta, one obtains the inverse cross section:

$$\sigma(\varepsilon_v) = \frac{\hbar \omega_{l=0} R_0^2}{2 \varepsilon_v} \cdot \ln\left( 1 + \exp\left( \pi \cdot \frac{\varepsilon_v - E_{l=0}}{\hbar \omega_{l=0}} - C(\varepsilon_v) \right) \right). \tag{18}$$

- *Energy just above the Coulomb barrier:*

When the energy is just above the barrier, Avishai's formulation reduces to Wong's prediction [41] in which the barrier is assumed to have the form of an inverted (full) parabola. The cross section is in this case not so much affected by the Coulomb slope and the calculation of the tunnelling only through the (full) parabola gives a satisfactory result:

$$\sigma(\varepsilon_v) = \frac{\hbar \omega_{l=0} R_0^2}{2 \varepsilon_v} \cdot \ln\left( 1 + \exp\left( 2\pi \cdot \frac{\varepsilon_v - E_{l=0}}{\hbar \omega_{l=0}} \right) \right). \tag{19}$$

Again, inclusion of expressions (18-19) into Eq. (1) would imply the use of numerical tools for solving the integral. To overcome this problem, in ABLA07 the effect of tunnelling on the particle decay width has been determined by fitting the numerical results of a complete calculation with the Avishei formula for the transmission coefficients: Firstly, the numerical solution of Eq. (1) is obtained without considering the tunnelling, resulting in the so-called classical decay width $\Gamma_{class}$. In the second step, Eq. (1) is integrated numerically with taking the tunnelling into account; this results in the so-called exact particle decay width $\Gamma_{exact}$. The ratio $\Gamma_{exact} / \Gamma_{class}$ is shown in Figure 2 for several different systems. This ratio is then fitted, and the obtained fitting function $f(E_f, A_f, A_v, V)$ is used in ABLA07, so that the exact solution of Eq. (1) can be approximated by $\Gamma_{ABLA} = f(E_f, A_f, A_v, V) \cdot \Gamma_{class}$.

The function $f(E_f, A_f, A_v, V)$ that fits best the ratio $\Gamma_{exact} / \Gamma_{class}$ has the following form:

$$f(E_f, A_f, A_v, V) = 10^{\left( 4 \cdot 10^{-4} \cdot x^{-\frac{4.3}{\ln(10)}} \right)} \quad \text{with} \quad x = \left( \frac{T}{(\hbar \omega)^2} \frac{1}{\mu^{1/4}} \right), \tag{20}$$

where $x$ is the ratio between the temperature ($T$) of the daughter nucleus ($A_f$) and the energy ($\hbar \omega$) of the inverse parabola at the potential barrier ($V$), divided by the forth root of the reduced mass ($\mu$) of the system. $\hbar \omega$ is calculated from the second derivative of the potential given in Section II.1.4.

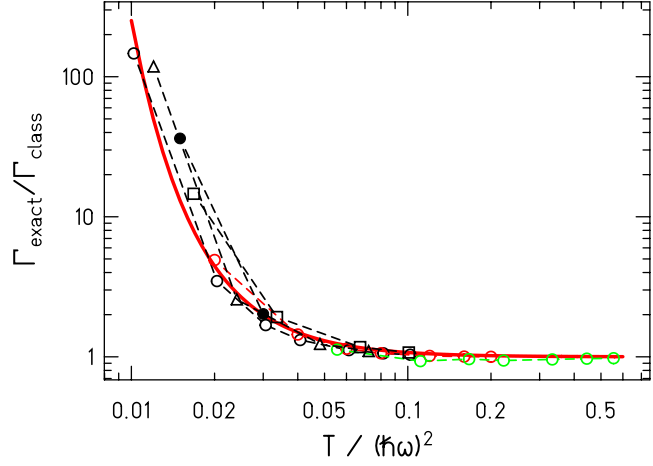

*Figure 2: Enhancement of the particle decay width due to tunnelling through the potential barrier for different particles and emitting systems (different symbols) together with a functional form given by Eq. (17) (full red line).*

### II.1.6. Expansion

In order to correctly describe the de-excitation of a heated nucleus, changes in the nuclear density of the compound nucleus with thermal energy have to be considered. A nucleus tends to expand when it is heated, until it reaches a status of thermal equilibrium, where the level density is maximal for the given total excitation energy. The increase of volume has three possible consequences which may affect the following de-excitation process: Firstly, it lowers the Coulomb barrier. Secondly, it changes the level density of the compound nucleus. Thirdly, and most important, the nucleus can enter the region of spinodal instabilities. In this section we discuss the first and second aspects, while the third aspect will be discussed in section III.

The radius of the expanded nucleus is derived from the analytical expression for the density at thermal equilibrium presented in Ref. [24]. There, the decrease of the density, $\rho_{eq}/\rho_o$, relatively to the normal density of the nucleus, is calculated according to the following formula:

$$\frac{\rho_{eq}}{\rho_o} = \frac{1}{4} \cdot \left(1 + \sqrt{9 - 8\frac{\varepsilon^*_{tot}}{\varepsilon_b}}\right), \tag{21}$$

where $\varepsilon^*_{tot} = E^*_{tot}/A$ is the excitation energy per nucleon ($E^*_{tot}$ is the total excitation energy of the system of mass number $A$) and $\varepsilon_b$ is the ground-state binding energy per nucleon of the system. Assuming a spherical nucleus, we obtain the relative increase of the radius, $r_{eq}/r_o$, due to thermal expansion:

$$\frac{r_{eq}}{r_o} = \left(\frac{1}{4} \cdot \left(1 + \sqrt{9 - 8\frac{\varepsilon^*_{tot}}{\varepsilon_b}}\right)\right)^{-\frac{1}{3}}, \tag{22}$$

The elongated nuclear radius at thermal equilibrium is used to calculate the nuclear potential (using the empirical formula of Bass [38]), and, finally the reduced Coulomb barrier.

The second effect of the thermal expansion is to change the level density. The level density is related to the thermal energy through the level density parameter, *a*. The dependence of *a* on the nuclear matter density is given by the Fermi-gas model:

$$\frac{a_{eq}}{a_o} = \left(\frac{r_{eq}}{r_o}\right)^2. \tag{23}$$

In Ref. [24] it is demonstrated that the above equation can apply for finite nuclei; specifically:

$$\frac{\tilde{a}_{eq}}{\tilde{a}_o} = \left(\frac{r_{eq}}{r_o}\right)^2. \tag{24}$$

In reality, we do not make use of the above formula (24) in ABLA07. Since the deexcitation cascade is ruled by the decay widths, which in turn depend on the relative weight of the level densities of the mother and daughter nuclei, the change on the density of levels due to thermal expansion will not reflect perceptibly on the decay widths. This is particularly true for heavy nuclei – where the difference in level density between mother and daughter is minimal – and at high excitation energy – where the density of levels is anyhow very high. For this reason, the effect of thermal expansion on the level density is not considered in ABLA07.

As an example of ideas described in the above sections, we show in Figure 3 a comparison between calculated and measured production cross sections of $^3$He and $^4$He in proton-induced reaction on $^{56}$Fe at several proton-beam energies. Calculations show only the contribution of the particle emission from the thermalised system, i.e. no production from the first stage of interaction (e.g. intra-nuclear cascade) is included.

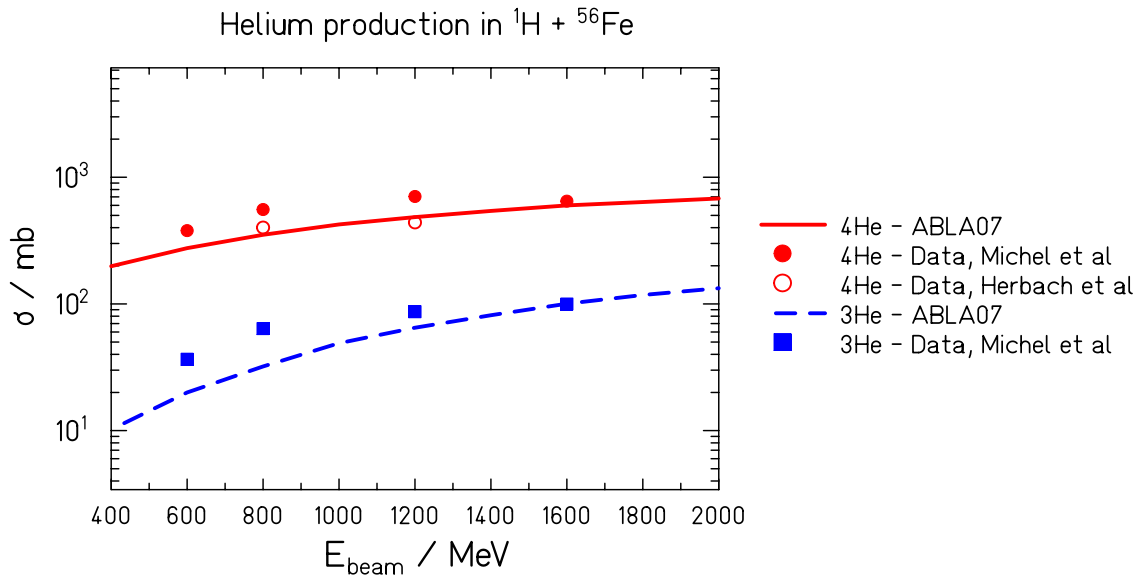

Figure 3: *Comparison between measured and calculated production cross sections of $^3$He and $^4$He – full ($^4$He) and dashed ($^3$He) lines: ABLA07 predictions; dots ($^4$He) and squares ($^3$He): data from Refs. [43,44]. Please note, that in the calculated cross sections no contribution from the first stage of the reaction is taken into account.*

## II.1.7. Kinetic-energy spectra

The kinetic energy of the emitted particle in the frame of the emitting source is sampled from the Maxwell-Boltzmann distribution at the corresponding temperature taking into account the effects of the Coulomb barrier for charged particles.

For generating random numbers following others than a rectangular function, different techniques are available. The option which is fastest in the application is based on the integration and the inversion of the function. In case of Maxwell distributions this procedure cannot be performed analytically, and usually one is performing this operation numerically, which has of course consequences on the computing time. To avoid this problem, we use in ABLA07 an appropriate random generator by a folding method. We demonstrate the procedure on the example of the Maxwellian energy distribution:

$$\frac{dI_{Maxwell}}{dE} \propto E \cdot e^{-E/T}. \tag{25}$$

The intensity $I$ is given as the product of the energy $E$ and an exponential decrease with a slope given by the temperature $T$.

A random generator for the exponential function

$$\frac{dI_{exp}}{dE} = e^{-E/T}$$

is readily given by:

$$E_i = -T \cdot \ln(PRN),$$

where $PRN$ is pseudo-random generator that produce numbers in the interval I={0,1} with uniform probability.

The Maxwell distribution can be obtained by the following folding expression of two exponential distributions:

$$\frac{dI_{Maxwell}}{dE} \propto \int_0^E e^{-\varepsilon/T} \cdot e^{-(E-\varepsilon)/T} d\varepsilon = e^{-E/T} \int_0^E d\varepsilon = E \cdot e^{-E/T}$$

Consequently, a random generator for the Maxwell distribution can be constructed by adding the results of two independent calls of the exponential random generator:

$$E_i = -T[\ln(PRN) + \ln(PRN)]. \tag{26}$$

In case of charged particles one has also to consider the influence of the Coulomb barrier. In this case, the Eq. (25) is read as:

$$\frac{dI_{Maxwell}}{dE} \propto \frac{E^2}{E+B} \cdot e^{-E/T}. \tag{27}$$

Due to the factor *(E+B)* in Eq. (27), one cannot obtain an exact formulation of the random number generator. In this case, we introduce an approximation: We start from the function:

$$dI/dE \propto E^2 \cdot e^{-E/T}, \tag{28}$$

for which one can obtain an exact formulation. The modification introduced by the additional Coulomb term *(B+x)* in Eq. (27) is small when *T<B*, and in this case Eqs. (27) and (28) are very close to each other. The difference between these two functions becomes more important for larger values of *T/B*, which is not often the case as the Coulomb barriers for light-charged particles are in most of the cases ~ 5 – 10 MeV.

For the function given by Eq. (28), according to the same ideas leading from Eq. (25) to Eq. (26) one obtains as the exact formulation the following expression:

$$E_i = -T(\ln(PRN) + \ln(PRN) + \ln(PRN)). \tag{29}$$

The same form can then be used for creating the spectra according to Eq. (27). Please note, that the logarithmic slope of the high-energy tail is correctly reproduced by this event generator.

In order to realistically calculate particle kinetic-energy spectra, functional forms given by Eqs. (25) and (27) have to be corrected for the quantum-mechanical effects at low particle kinetic energies, which lead to an additional factor proportional to $1/\upsilon$, where $\upsilon$ is the particle velocity [45]. In this case, Eqs. (25) and (27) have the following forms:

$$\frac{dI}{dE} \propto \begin{cases} \sqrt{E} \cdot \exp\left(-\dfrac{E}{T}\right), & \text{neutrons} \\ \dfrac{E^{3/2}}{E+B} \cdot \exp\left(-\dfrac{E}{T}\right), & \text{charged particles} \end{cases} \tag{30}$$

For these two functions, one cannot get exact formulations of the random-number generator, but similar as in case of Eq. (27) an approximation, which enables fast calculations of kinetic energy spectra:

$$E_i = \begin{cases} 2 \cdot T \cdot \sqrt{\ln(PRN) \cdot \ln(PRN)}, & \text{neutrons} \\ 3 \cdot T \cdot (-\ln(PRN) \cdot \ln(PRN) \cdot \ln(PRN))^{1/3}, & \text{charged particles} \end{cases} \tag{31}$$

Equation (31) is then used for obtaining the kinetic energies of emitted particles. In Figure 4, a comparison between neutron and proton kinetic-energy spectra calculated according to Eqs. (30) and (31) is shown.

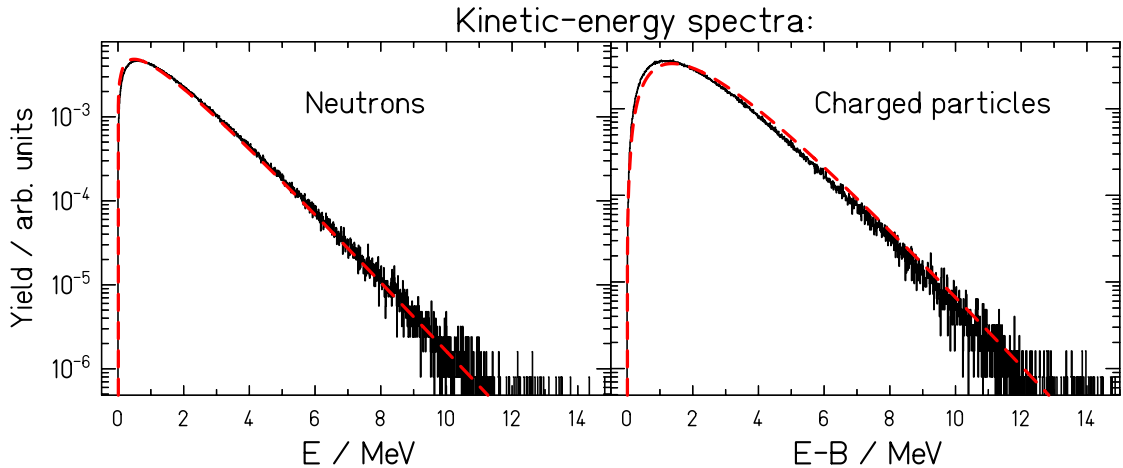

*Figure 4: Comparison between analytical functions specified by Eq. (30) (dashed line) and the corresponding random generators specified by Eq. (31) (full histogram). The parameters are T = 1 MeV and B = 10 MeV.*

After determining the kinetic energy of the emitted particle, its velocity vector is determined assuming isotropic emission[3] in the frame of the emitting source. Using this information, at every de-excitation step the recoil of the excited nucleus due to particle emission is then calculated.

## II.2. Gamma emission

In several evaporation codes, γ-radiation is not included as a possible channel, because the particle decay channels dominate above the particle-emission threshold. However, in the last de-excitation step of the evaporation cascade, gamma emission becomes competitive to particle decay for heavy compound nuclei. Normally, the emission of gammas is much less probable than the particle decay (about $10^5$ times less favourable). Since the level density depends on the mass (heavier nuclei have denser energy levels) the number of levels between the ground state and the particle separation energy of a heavy nucleus can be as high as $10^5$ or even exceed this value. If the excitation energy of the compound nucleus is slightly higher than its particle separation energy, it can decay only into the ground state or into the first excited states of the daughter nucleus (if the daughter nucleus is an even-even nucleus, then only the ground state is energetically accessible due to the pairing gap – see Ref. [46] for a wider discussion). In this situation, gamma emission and particle decay can become two competitive channels.

As the emission of statistical γ-rays occurs predominantly via the giant dipole resonance, the γ-radiation rate can be calculated according to Ref. [36] as:

$$\Gamma_\gamma(E) = \sum_{I=|J-1|}^{J+1} \int_0^E \varepsilon_\gamma^3 \cdot k(\varepsilon_\gamma) \cdot \frac{\rho(E - \varepsilon_\gamma, I)}{\rho(E, I)} d\varepsilon_\gamma \ , \qquad (32)$$

where $E$ is the excitation energy of the mother nucleus and $k(\varepsilon_\gamma)$ is the radiative strength function for a dipole electric transition. As already said, for high excitation energy the probability for γ emission is negligible compared to the probability for particle emission and it becomes important only at the energies around and below the particle separation energies. As indicated in Ref. [36], taking $E = S_n$, and using the power approximations for the radiative strength function [47] and the constant temperature model [36], equation (32) can be parameterised as:

$$\Gamma_\gamma(S_n) = 0.624 \cdot 10^{-9} \cdot A^{1.60} \cdot T^5 \text{ MeV}, \qquad (33)$$

where $A$ is the mass of a mother nucleus and $T$ is the nuclear-temperature parameter of the constant-temperature model [36].

The effects of gamma decay are especially visible in the strength of the even-odd staggering of the final products [46]. As an example, the production cross sections of different isotopes of $_{71}$Lu in the reaction $^{208}$Pb (1$A$ GeV) + $^1$H are shown in Figure 5. The experimental data from [48] are shown as full dots and compared with two sets of calculation: without including γ emission (open squares) and with including it (open triangles). One can observe that the γ competition tends to reduce the even-odd structure in the isotope cross sections to a great extent.

## II.3. Fission

Fission plays an important role in the decay of heavy nuclei. At each de-excitation step a competition between fission and other decay channels is calculated. The fission decay width is calculated in a time-dependent approach as developed in Refs. [21,25,26]. If fission occurs, the ABLA07 code calls a program called PROFI where masses, atomic charges, excitation energies and velocities of two fission fragments are calculated. In the PROFI code, only binary fission is considered. The original version of the PROFI model has been published in Refs. [49,50]; recent developments and improvements are given in Refs. [51,52].

---

[3] This approximation is valid for moderate angular momentum or high excitation energies.

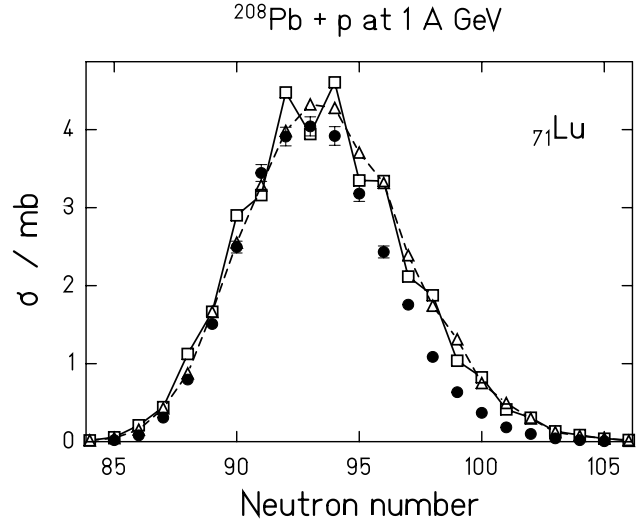

*Figure 5: Production cross sections of the isotopes of lutetium produced in the reaction $^{208}Pb+H$ at 1 A GeV, calculated with the statistical code ABRABLA with (Δ) and without (□) the inclusion of the γ-radiation decay channel, and compared to the experimental data (●) from Ref. [48]. The errors on the experimental data are shown only if the error bars are larger than the symbol size.*

### II.3.1. Time-dependent fission width

The modelling of the fission decay width at high excitation energies requires the treatment of the evolution of the fission degree of freedom as a diffusion process, determined by the interaction of the fission collective degree of freedom with the heat bath formed by the individual nucleons [53,4]. Such process can be described by the Fokker-Planck equation (FPE) [54], where the variable is the time-dependent probability distribution $W(x, p; t, \beta)$ as a function of the deformation in fission direction $x$ and its canonically conjugate momentum $p$. The parameter $\beta$ is the reduced dissipation coefficient. The solution of the FPE leads to a time-dependent fission width $\Gamma_f(t)$. However, these numerical calculations are too much time consuming to be used in nuclear-reaction codes.

To avoid this problem, an analytical approximation to the solution of the one-dimensional Fokker-Planck equation for the time-dependent fission-decay width for the initial condition of a Gaussian distribution centred at the spherical shape has been developed in Refs. [25,26]. The mean values and the widths of the initial Gaussian distributions in space and momentum are given by the entrance channel. In this approximation, the time dependence of the fission width is expressed as [25,26]:

$$\Gamma_f(t) = K \cdot \Gamma_{BW} \cdot \frac{W_n(x=x_b;t,\beta)}{W_n(x=x_b;t\to\infty,\beta)}, \qquad (34)$$

where $K = \left[1+(\beta/(2\omega_0))^2\right]^{1/2} - \beta/(2\omega_0)$ is the Kramers factor [53] with $\omega_0$ corresponding to the frequency of the harmonic oscillator describing the potential at the saddle-point deformation and $\beta$ is the reduced dissipation coefficient. In the above equation, $\Gamma_{BW}$ is the fission width given by the statistical approach of Bohr and Wheeler [55] and $W_n(x=x_b;t,\beta)$ is the normalized probability distribution at the saddle-point deformation $x_b$. The saddle-point deformations are calculated according to Ref. [56].

In case of a nuclear potential approximated by a parabola, the solution of the Fokker-Planck equation for the probability distribution $W(x=x_b;t,\beta)$ at the saddle-point deformation has a Gaussian form with a time-dependent width. For a special case of initial conditions, namely zero mean deformation and zero mean velocity, this solution has the following form [57]:

$$W_n(x = x_b; t, \beta) = \frac{1}{\sqrt{2\pi}\sigma} \cdot \exp\left(-\frac{x_b^2}{2\sigma^2}\right), \qquad (35)$$

with $\sigma^2$ given as [57]:

$$\sigma^2 = \frac{kT}{\mu\omega_1^2}\left\{1 - \exp(-\beta \cdot t) \cdot \left[\frac{2\beta^2}{\beta_1^2}\sinh^2\left(\frac{1}{2}\beta_1 t\right) + \frac{\beta}{\beta_1}\sinh(\beta_1 t) + 1\right]\right\}, \qquad (36)$$

where $k$ is Boltzmann's constant, $T$ is the nuclear temperature, $\mu$ is the reduced mass associated to the deformation degree of freedom, $\omega_1$ describes the curvature of the potential at the ground state and $\beta_1 = (\beta^2 - 4\omega_1^2)^{1/2}$.

Due to the classical nature of the Fokker-Planck equation, the initial behaviour predicted by this solution is wrong since for $t = 0$ equation (36) leads to $\sigma = 0$. Therefore, in Refs. [25,26] the zero-point motion at the spherical shape has been chosen as the initial condition of the problem. The zero-point motion is taken into account by shifting the time scale $t \to t + t_0$ in Eq. (36) by a certain amount $t_0$, where $t_0$ is the time needed for the probability distribution to reach the width of the zero-point motion in deformation space. The value of $t_0$ is calculated as [21]:

$$t_0 = \frac{1}{\beta}\ln\left(\frac{2T}{2T - \hbar\omega_1}\right), \text{ in the under-damped regime } (\beta < 2\omega_1), \text{ and}$$

(37)

$$t_0 = \frac{\hbar\beta}{4\omega_1 T}, \text{ in the over-dumped regime } (\beta \geq 2\omega_1).$$

In Figure 6 a comparison between the numerical solution of the one-dimensional Langevin equation of motion (full histogram) and the analytical approximation for case of $^{248}$Cf starting from spherical initial conditions is shown. The agreement between these two solutions is very satisfactory. For more details, see [25,26].

By introducing the time-dependent fission decay width, the ABLA07 code can be considered as a dynamical code with the explicit treatment of the system time evolution. Technical details on the inclusion of the time evolution are given in the Appendix 1 of Ref. [26].

### II.3.2. Influence of initial conditions

In the previous section we gave a brief overview on the analytical approximation of the time-dependent fission width developed in Refs. [25,26] for spherical initial conditions. On the other hand, it is very difficult to create a fissioning system under such ideal initial conditions [58], and, therefore, the influence of initial deformation on the fission decay width has to be taken into account in order to have a realistic description of this decay channel [58, 59,60]. In Ref. [59], we extended the above-described approach, which has been derived for the initial condition of a Gaussian distribution centred at the spherical shape, to more general initial conditions; here, a short overview will be given.

In order to take into account non-spherical initial conditions, we introduced into the approximation (34) - (37) the solution of the dynamic Langevin equation of the system without considering the fluctuating term, assuming that the system starts at the finite initial deformation $x_{init}$. For this case, we calculate the mean deformation of the system at each time *t*. We get two solutions, one for the over-damped and one for the under-damped regime.

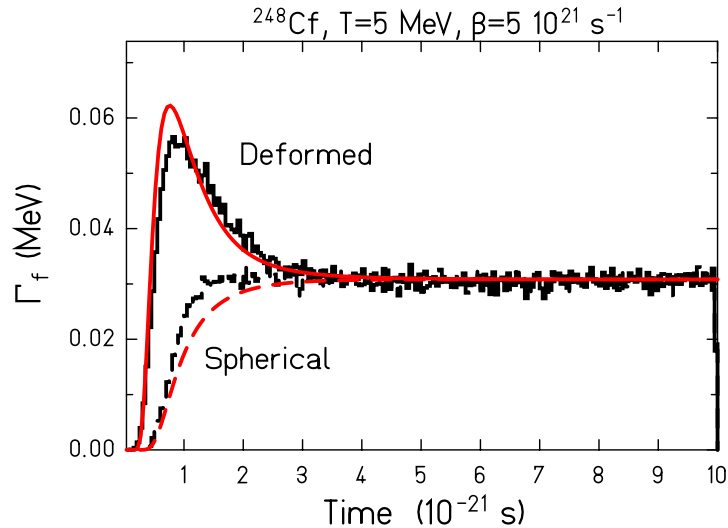

*Figure 6: Time-dependent fission decay width $\Gamma_f(t)$ as obtained from the solution of the one-dimensional Langevin equation of motion (histograms) assuming that the excited $^{248}$Cf system starts from either a spherical ($\beta_2=0$) or a deformed ($\beta_2=0.235$) configuration. Dashed and full lines correspond to the result obtained with the described analytical approximations for spherical [25,26] and deformed [59] initial conditions, respectively. The figure shows the case T=5 MeV and $\beta = 5 \cdot 10^{21} s^{-1}$.*

In the over-damped regime, the mean deformation of the system at time *t* follows the equation:

$$x_{mean} = x_{init} \cdot \exp\left[-\frac{1}{2}(\beta - \beta_1) \cdot t\right]. \tag{38}$$

In the under-damped regime, the mean deformation of the system is described by the equation:

$$x_{mean} = x_{init} \cdot \cos\left(\frac{1}{2}\beta_2 \cdot t\right) \cdot \exp(-\beta \cdot t), \text{ with } \beta_2 = \sqrt{4\omega_0^2 - \beta^2}. \tag{39}$$

The solutions (38) and (39) are then included into the solution of the Fokker-Planck equation given in (34) by performing the transformation $x_b \to x_b - x_{mean}$.

This then leads to the following analytical approximation to the solution of the Fokker-Planck equation for the time-dependent fission width:

$$\Gamma_f(t) = K \cdot \Gamma_{BW} \cdot \frac{kT}{\mu\omega_1^2 \cdot \sigma^2} \cdot \frac{\exp\left(-\frac{(x_b - x_{mean})^2}{2\sigma^2}\right)}{\exp\left(-\frac{\mu\omega_1^2 \cdot x_b^2}{2kT}\right)},$$

(40)

$$\sigma^2 = \frac{kT}{\mu\omega_1^2}\left\{1-\exp(-\beta\cdot(t+t_0))\cdot\left[\frac{2\beta^2}{\beta_1^2}\sinh^2\left(\frac{1}{2}\beta_1(t+t_0)\right)+\frac{\beta}{\beta_1}\sinh(\beta_1(t+t_0))+1\right]\right\},$$

where $t_0$ is given by Eq. (37) and $x_{mean}$ by Eqs. (38) and (39). This is the formula used in ABLA07 to calculate the fission decay width.

In Figure 6, we compare the results of this analytical approximation for the time-dependent fission width with the numerical results of the Fokker-Planck equation calculated for the nucleus $^{248}$Cf starting from deformed initial conditions (full line and full histogram). The agreement between the analytical approximation for more general initial conditions and one-dimensional numerical calculations is quite satisfactory.

### II.3.3. Low-energy fission

In case of low-energy fission, the double-humped structure in the fission barrier as a function of elongation and the symmetry classes at different saddle points are of importance for a proper description of the process. These effects have been included in the ABLA07 code, following the ideas developed in Refs. [1,61,62]: Assuming that the vibrational states in the second well are completely damped into all the other compound states, i.e. the system found in the second minimum can either fission via passage over the second (B) barrier or return to the initial deformation via passage over the first (A) barrier, the fission decay width can be calculated as [1,61]:

$$\Gamma_f = \frac{\Gamma_A \cdot \Gamma_B}{\Gamma_A + \Gamma_B}, \tag{41}$$

where, $\Gamma_A$ and $\Gamma_B$ represent the partial decay widths for fission over barrier A and B, respectively. These partial widths are calculated as:

$$\Gamma_{A,B} = \frac{1}{2\pi\cdot\rho_g(E)}\int_0^{E-B_f^{A,B}}\rho_{A,B}(\varepsilon)d\varepsilon. \tag{42}$$

In the above equation, $\rho_g$ is the level density at the initial deformation, $\rho_{A,B}$ level density above the barrier A and B, respectively, and $B_f^{A,B}$ the height of the barrier A and B, respectively.

In order to calculate the level density at a specific deformation, one has to take into account the symmetry class of the corresponding configuration. Following the ideas of Refs. [1,62,63] we assume that the barrier A is mirror symmetric and axially symmetric for nuclei with $N\leq 144$, while axially asymmetric for nuclei with $N>144$. The barrier B is axially asymmetric, and mirror symmetric for nuclei with mass smaller than 226, while mirror asymmetric for larger masses. We also assume, that for nuclei with $Z^2/A$ less than 34 only barrier B plays a role, while for heavy nuclei with $Z^2/A$ larger than 40.6 only barrier A is important. In the intermediate region both barriers have to be considered.

Another important input in the Eq. (42) is the height of the corresponding barrier. In ABLA07 we assume that these two barriers have the same height and that it is given by the prediction of the finite-range liquid drop model of Sierk [64] with ground-state shell-correction energies of Ref. [29] included. We make this assumption for the following two reasons: Firstly, experimental information on the fission-barrier height is available for a very limited number of nuclei (see e.g. [65]), with large uncertainties for the barrier which is the lowest between the two A and B. Moreover, different theoretical calculations predict often very different values of the barrier heights, and sometimes they over/under-predict the experimental barrier by few MeV. This all makes it quite difficult, or even impossible, to perform the calculations in regions where the experimental data are scarce or even not existing. Secondly, several studies [30,31,66] have shown that the shell-correction energy at the fission saddle point is very small, and, thus, considering the uncertainties in model predictions, can be neglected. Due to all this, we assume that $B_A = B_B = B_{FRLDM} - \delta U_{GS}$, where $\delta U_{GS}$ is ground-state shell correction energy, and $B_{FRLDM}$ macroscopic fission barrier calculated according to Ref. [64]. We have decided to use the model of Ref. [64] as according to a recent study performed in [67], this model gives very realistic predictions of fission-barrier heights in experimentally unexplored regions.

In Figure 7, we compare the prediction of ABLA07 with measured fission probability as a function of excitation energy for the compound system $^{235}$Np. The agreement between the data and calculations is very satisfactory.

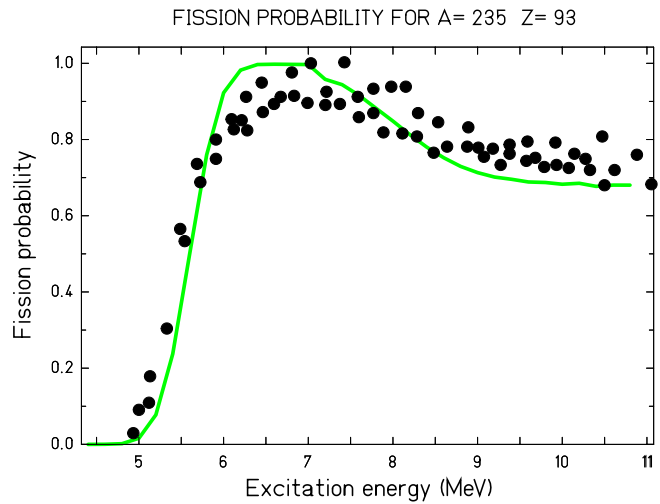

*Figure 7: Energy-dependent fission probability for the compound system $^{235}$Np: full symbols – experimental data from Ref. [1], full line – results of ABLA07.*

### II.3.4. Fragment production in fission

Properties of fission fragments, i.e. masses, atomic numbers, excitation and kinetic energies, are calculated based on the macro-microscopic approach and the separability of compound-nucleus and fragment properties on the fission path [51]. The original technical description of the fragment-formation model – PROFI – was published in Refs. [49,50], while the updated description will be the subject of a forthcoming publication.

In the PROFI model it is assumed that different splits in mass are basically determined by the number of available transition states above the potential energy surface behind the outer saddle point. The macroscopic properties of the potential-energy landscape of the fissioning system are attributed to the strongly deformed fissioning system, which are deduced from mass distributions at high excitation energy [68] and Langevin calculations [69]. The microscopic properties of the potential-energy landscape of the fissioning system are given by the qualitative features of the shell structure in the nascent fragments. They are determined from the observed features of the fission channels [70] according to the procedure described in [51].

In case of spontaneous fission, the mass distribution is not determined by the phase space but by the variation of the tunnelling probability through the outer barrier as a function of mass asymmetry. The tunnelling probability is calculated using the Hill-Wheeler approach.

The dynamics of the fission process responsible for the fragment formation is considered in an approximate way: Since a variation of the mass asymmetry is connected with a substantial transport of nucleons and, consequently, the inertia of this collective degree of freedom should be large, we assume that the phase space near the outer saddle point determines the mass asymmetry of the system, which is more or less frozen during the descent to scission. On the other hand, the *N/Z* collective degree of freedom can be considered as a fast degree of freedom, as it is enough to exchange very few neutrons or protons between the two nascent fragments in order to explore the full *N/Z* range observed in the final fragments. Therefore, we assume that the *N/Z* degree of freedom is determined, opposite to mass asymmetry, near the scission point, and we calculate its value taking into account the charge-polarisation effects [71].

The excitation energies of the created fragment are calculated from the available excitation energy at the scission point and the deformation energies of the fragments at scission. The deformation energies of the fragments are assumed to be specific to the individual fission channels. They are deduced from experimental data on total kinetic energies and neutron yields. Kinetic energies are then calculated applying the energy conservation law.

### II.3.5. Particle emission in fission

In ABLA07, particle emission is calculated at different stages of the fission process – (i) up to the saddle point, (ii) from the saddle up to the scission point, and (iii) from the two separated fission fragments. In order to calculate the particle emission on the way from the saddle to the scission point, we have parameterized the saddle-to-scission times obtained by solving the three-dimensional Langevin equation of motion using the one-body dissipation tensor with the reduction coefficient $K_s=0.25$ [72]. Then, at each time step, the probability to emit a neutron or some of the light charged particles is calculated. IMF emission is not considered as a decay channel between saddle and scission. This procedure is repeated as long as the cumulative particle emission time (i.e. sum of the particle emission times emitted after the saddle point) is shorter than the saddle-to-scission time.

After scission, two fission fragments are formed, and their decay is followed as described in Section II.1.

In Figure 8 a comparison between measured and calculated fission-fragment mass and neutron-multiplicity distributions in case of spontaneous fission of $^{252}$Cf is shown. Please note that there was no special adjustment of model parameters in order to reproduce the data.

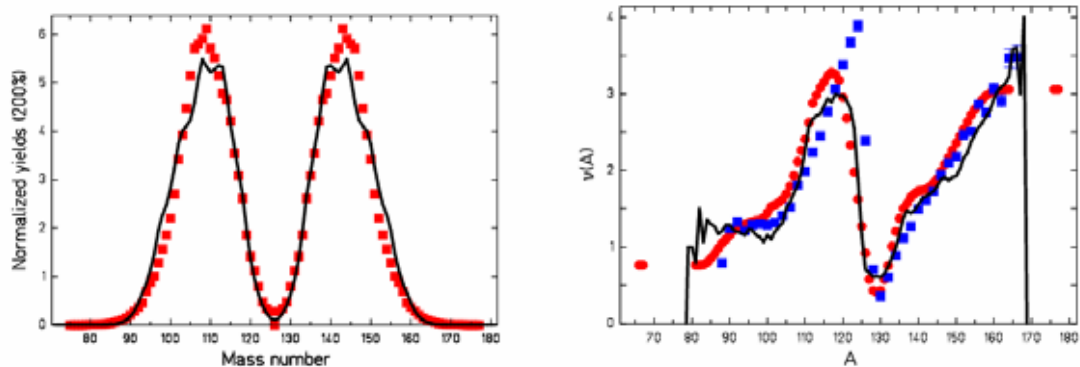

*Figure 8: Spontaneous fission of $^{252}$Cf – Left: comparison between measured mass distribution [73] (symbols) and ABLA07 prediction (full line); Right: comparison between measured [74] (dots) and evaluated [75] (squares) neutron multiplicities as a function of the fission-fragment mass and the result of an ABL07 calculation (full line).*

### II.4. IMF emission

The range of emitted fragments in the ABLA07 code has been extended to above $Z = 2$ in order to obtain a more realistic description of the production of intermediate-mass fragments (IMFs), which was strongly underestimated in the previous version of ABLA. Two models for the production of IMFs are implemented: In the first scenario, all nuclei below the Businaro-Gallone maximum of the mass-asymmetry dependent barrier, see Figure 9, are taken into account in the evaporation process. The barriers are given by the Bass nuclear potential. Thermal expansion of the compound nucleus is considered. In the second scenario, which will be described in Section III, if the excitation energy of the system exceeds the corresponding threshold, the simultaneous break-up of the system is modelled according to a power-law distribution, which is suggested by several theoretical models.

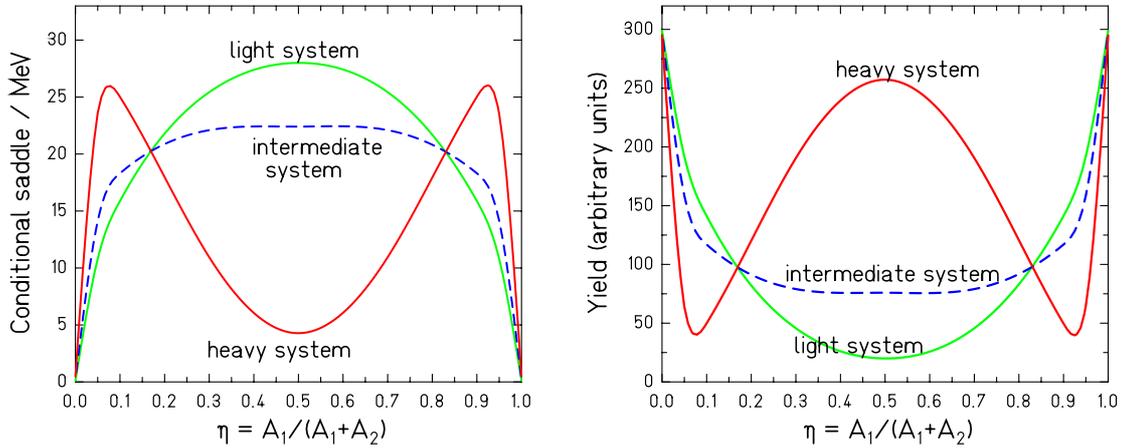

*Figure 9*: *Energies above the ground state in the touching-sphere configuration (left) and corresponding mass distributions (right) given by the available phase space above corresponding configuration in the left part of the figure.*

In the case of sequential IMF emission, in order to have a fast calculation scheme, the different decay channels are divided into a few groups: The emission of neutrons, light charged particles and gammas is treated explicitly. The same is true for fission. The emission of IMFs with $Z \geq 3$, on the other hand, is treated as one class of events in the first step, in order not to increase the computational time. The idea is the following:

To calculate the probability $P_i$ of a given decay channel $i$, we need the corresponding decay width $\Gamma$:

$$P_i = \frac{\Gamma_i}{\Gamma_{tot}}, \quad \Gamma_{tot} = \sum_k \Gamma_k = \Gamma_{neutron} + \sum_{lcp} \Gamma + \Gamma_{gamma} + \Gamma_{fission} + \sum \Gamma_{IMF}, \quad (43)$$

In the above equation, the sum over *lcp* goes over all light-charged particles with $Z=1, 2$, while the sum over IMF goes over all intermediate-mass fragments that can be emitted in a given reaction. Therefore, the explicit calculation of the last term in Eq. (43) would be very time consuming. On the other hand, from the experimental observations we know that the element distribution of IMF fragments follows a power law. Thus, we can well estimate the total decay width for IMF production ($\Gamma_{IMF}^{tot} = \sum \Gamma_{IMF}$) by determining the slope in the double-logarithmic presentation by calculating the decay width for the isotopes of two elements (e.g. $Z = 3$ and $Z = 5$) and integrating the adapted power-law function:

$$\Gamma_{IMF}(Z) = a \cdot Z^b \Leftrightarrow \log(\Gamma_{IMF}(Z)) = a + b \cdot \log(Z) \quad \Rightarrow$$

$$b = \frac{\log(\Gamma_{IMF}(Z=5)/\Gamma_{IMF}(Z=3))}{\log(5/3)}, \quad a = \frac{\Gamma_{IMF}(Z=5)}{5^b}$$

$$\Gamma_{IMF}(Z=i) = \sum_A \Gamma_{IMF}(Z=i,A), \quad i=3,5, \tag{44}$$

where $A$ is the mass of a selected IMF. $\Gamma_{IMF}(Z=3)$ and $\Gamma_{IMF}(Z=5)$ are then explicitly calculated according to the procedure described below.

Once the parameters $a$ and $b$ are obtained, one can determine the total decay width for IMF emission by performing the following integration:

$$\Gamma_{IMF}^{tot} = \int_3^{Z_{CN}} a \cdot Z^b dZ = a \cdot \left(Z_{CN}^{b+1} - 3^{b+1}\right). \tag{45}$$

Only if the emission of IMFs is realised, the competition between the individual IMFs is to be considered, as described below.

Since long time, it has been discussed whether the emission of an IMF from a heavy nucleus (above the Businaro-Gallone point) is better described as an evaporation process or as a fission process with very asymmetric mass-split. Both approaches were already used in the past in nuclear de-excitation codes, e.g. in GEMINI [76] as very asymmetric fission or in GEM2 [77] as evaporation. Already in 1975 it was pointed out that there is a continuous transition between the two processes [78]. Recently [79] it was shown that even for such a heavy nucleus as $^{238}$U the lightest IMFs are produced in a rather compact configuration, indicating that there is gradual transition from the standard fission process towards evaporation. From the physical point of view an extremely asymmetric binary split into two compact nuclei corresponds to an evaporation of a light nucleus from a heavy compound nucleus. In ABLA07 we based the fission-to-evaporation changeover on the M-shaped potential energy as a function of the mass asymmetry. At the point were the M-shaped potential reaches it maximum, the fission model smoothly fades away in favour of the evaporation process.

In ABLA07, the statistical weight for the emission of IMFs is calculated, similarly as in case of any other particle-decay channel, on the basis of the detailed-balance principle, except that in this case also the available nuclear levels in the IMF have to be considered The decay width ($\Gamma$) as a function of the excitation energy ($E$) depends on the inverse cross section ($\sigma_{inv}$), on the level densities of the two daughter nuclei ($\rho_{imf}$ and $\rho_{partner}$) and on the level density of the mother nucleus above the ground state ($\rho_C$):

$$\Gamma \approx \int_0^{E_{imf}^{max}} \int_0^{E_{partner}^{max}} \sigma_{inv} \frac{\rho_{imf}(E_{imf}) \cdot \rho_{partner}(E_{partner})}{\rho_C(E)} dE_{imf} \, dE_{partner}, \tag{46}$$

with the following relation that guaranties the energy conservation:

$$E = E_{imf} + E_{partner} + Q + \varepsilon. \tag{47}$$

Here $E$, $E_{imf}$ and $E_{partner}$ represent the initial excitation energy of the mother nucleus, and the excitation energies of the two daughter nuclei, respectively. Q is the Q-value, and $\varepsilon$ is the total kinetic energy in the centre of mass of the system. The barrier ($B$) which is also playing the role is calculated using the fusion nuclear potential of Bass [38] (see also Section II.1.4). The inverse cross section ($\sigma_{inv}$) is calculated using the ingoing-wave boundary condition model [22], where only the real potential is used to describe the transmission probability of particles. An analytical approximation to Equation (46) is used in order to avoid the numerical calculation of the two integrals, which is rather time-consuming: We assume that in order to calculate the phase space available for the IMF emission, we can, instead of folding the level densities (Eq. (46)) of the two fragments at the saddle point for IMF emission, calculate the level density of the compound nucleus at the same intrinsic excitation energy, using a modified level-density parameter to consider the increased surface of the configuration at the barrier In other words, instead, as described by Eq. (46), considering the system in a moment of the IMF emission as two systems (IMF and its partner) in the touching-sphere configuration, we describe it as a single system in the given configuration, i.e. deformation, angular momentum and excitation energy given by the touching-sphere configuration. To test this assumption, we have calculated the decay width for the $^{16}$O emission from several different compound systems, using either the Eq. (46) or assumption of a single system in the touching-sphere configuration, resulting in $\Gamma_2$ and $\Gamma_1$, respectively. In Figure 10, we show the ratio between $\Gamma_2$ and $\Gamma_1$ as a function of excitation energy above the touching-sphere configuration.

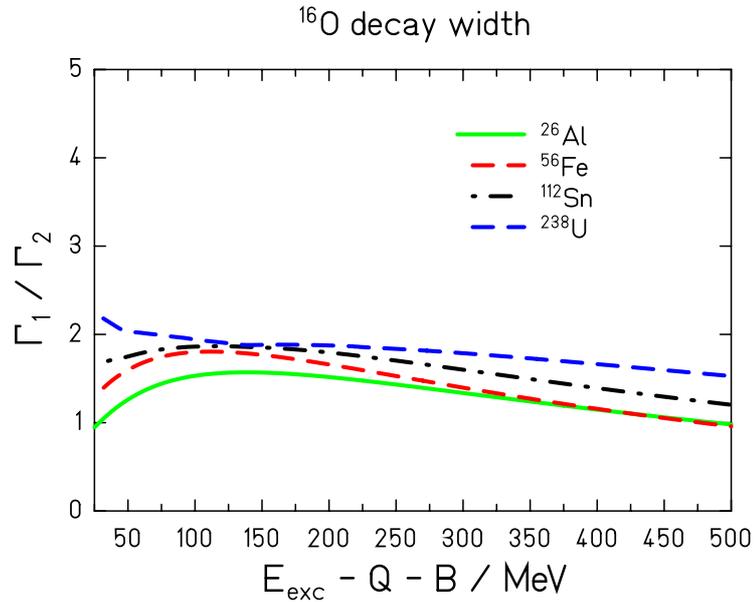

*Figure 10: Ratio between $^{16}$O decay widths, $\Gamma_1$ and $\Gamma_2$, calculated assuming one system in the touching-sphere configuration or two systems ($^{16}$O and its partner) in the same configuration, respectively.*

The kinetic energies of sequentially emitted IMFs and their partners are calculated, similar as in case of fission, from Coulomb repulsion using the momentum conservation in the frame of the decaying mother nucleus.

In Figure 11 experimental data measured in the reaction $^{238}$U+$^{1}$H at 1 $A$ GeV [79,80,81,82] are compared with the predictions of ABLA07 coupled to the reaction model BURST [79]. In this reaction, the largest contribution to the production of residual nuclei is coming from fission. On the other hand, nuclei with atomic number smaller than ~15 are produced as intermediate-mass fragments, while those with atomic number larger than ~70 are residues after the sequential emission of neutrons, light-charged particles and/or IMFs. Cross sections, as well as first and second moments of the isotopic distributions are compared, and agreement between the data and the calculations is very satisfactory.

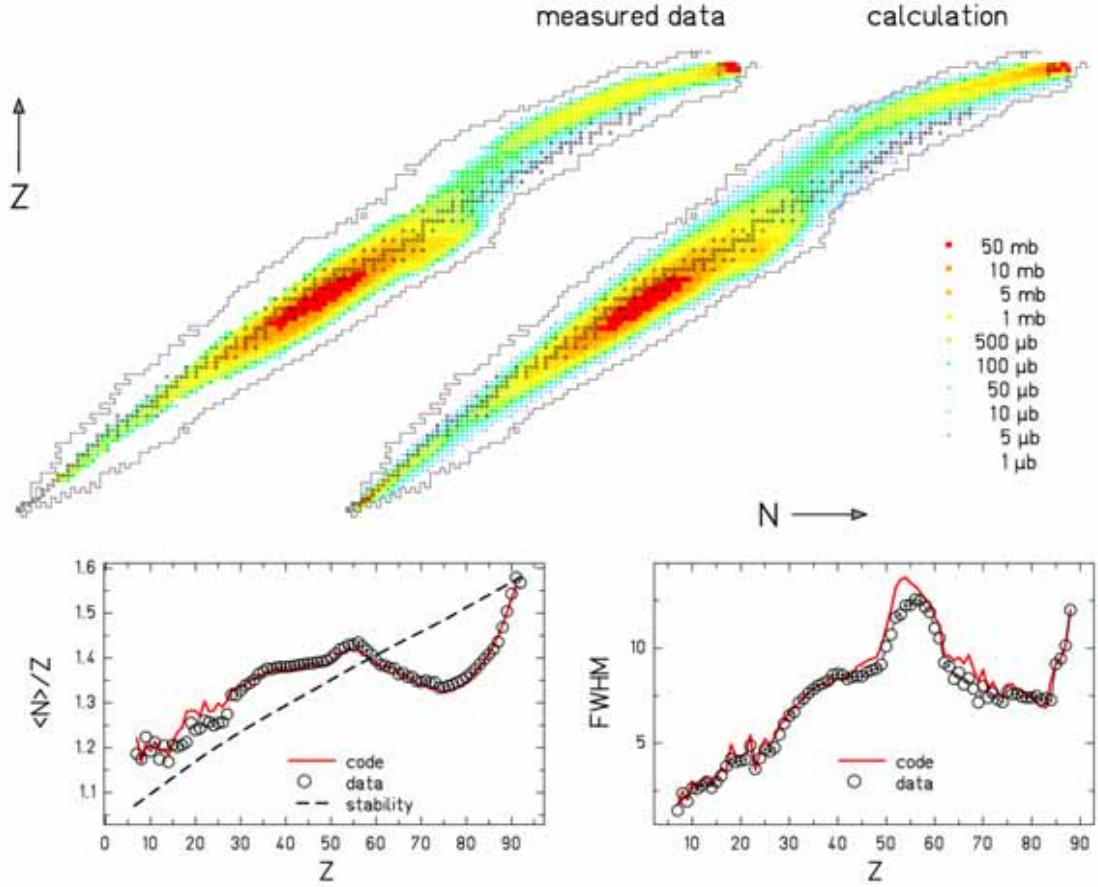

*Figure 11: Up – Cross sections for the nuclei produced in 1 GeV p on $^{238}$U: Measured cross sections [79,80,81,82] (left) and prediction of BUSRT [79] + ABLA07 (right) presented on the chart of the nuclides. Down – Left: Mean neutron-to-proton ratio of isotopic distributions as a function of the atomic number, compared with the stability line (dashed line) and to the BURST + ABLA07 prediction (solid line). Right: FWHM of the isotopic distributions compared to the prediction of the BURST + ABLA07 code (solid line).*

## III. Break-up stage

If the excitation energy acquired during the first, collision, stage is high enough, the increase of volume has a dramatic consequence: The nucleus enters the spinodal region [83] characterized by negative incompressibility. In this region, an increase in the system volume due to expansion is connected with the increase in pressure, and, consequently, any local fluctuation in density is strongly amplified leading to a mixed phase consisting of droplets represented by a small amount of light nuclei at normal nuclear density, and the nuclear gas represented by individual nucleons. This process is often called "break-up". The fragments formed in this process undergo deexcitation process and cool down. What is finally experimentally observed are the cold fragments, normally called IMFs. The entire multifragmentation process is scientifically very interesting for its relation to the equation-of-state of nuclear matter, in particular to the liquid-gas phase transition.

The starting point of the break-up stage in ABLA07 is a hot nuclear system –so-called "spectator"[4], leftover of the initial collision stage. We assume that, if the excitation energy per nucleon of the spectator exceeds a limiting value [9], the system undergoes the break-up stage; otherwise we assume that it will directly de-excite through sequential evaporation and/or fission.

---

[4] The term spectator is derived from fragmentation reactions, but the following description of the break-up process is valid for the decay of any hot thermalised system regardless of the way how it was produced.

About the limiting excitation energy per nucleon, two options are possible in ABLA07. The default option is that the limiting excitation energy per nucleon is constant for all nuclei; its value is fixed to 4.2 MeV. Another possible option is to use a mass-dependent value of the limiting excitation energy, deduced from the mass dependence of the temperature in the plateau of the caloric curve as pointed out by Natowitz in [84].

Please note that in the description of the break-up stage we do not consider any effect of compression, which could play a role in case of central heavy-ion collisions at Fermi energies. In case of nucleus-nucleus collisions at relativistic energies or of spallation reactions, the heating of the system is purely thermal without any influence of compression; for these reactions, the break-up stage in ABLA07 is adapted.

### III.1. IMF formation by break-up

It is not trivial to determine theoretically the size distribution of the break-up fragments. Models that evaluate it just by phase-space arguments, considering all possible partitions and weighted them by the number of available states, are considered to be inadequate since they neglect the dynamic of the expansion. On the other hand, the dynamics of the break-up process is far to be fully understood. In this context, in order to have an estimate of the production cross-sections of the IMFs, we based our model on the following considerations:

At the starting point of ABLA07, the spectator nucleus has mass $A_{init}^{spectator}$ and excitation energy $E_{init}^{spectator} = A_{init}^{spectator} \cdot \varepsilon_{init}$. If $\varepsilon_{init}$ is larger than some limiting value $\varepsilon_{freeze\_out}$ [9], the system will enter the break-up stage, where the excitation energy of the spectator is partially consumed to break up the spectator into several hot fragments. In the light of this picture, the break-up process in ABLA07 is technically divided into two steps.

As the first step, it is calculated how much of the initial energy is removed through the loss of mass to form nucleons or fragments (which are, at this stage, not specified). Specifically, it is calculated to which amount the mass of the spectator has to be reduced, down to $A_{freeze-out}^{spectator}$, in order to get to an excitation energy per nucleon corresponding to $\varepsilon_{freeze\_out}$. The energy consumed to lose one mass unit varies from 10 MeV for an initial excitation energy of 2.9 $A$ MeV to 5 MeV for an initial excitation energy of 11.8 $A$ MeV. These values have been deduced from the comparison with the experimental data in the reaction $^{238}$U+Pb at 1 $A$ GeV [85]. In the model, we assume that in the break-up stage the N/Z ratio is conserved, so the break-up product has the same N/Z ratio as the initial spectator nucleus. In this way, we obtain the mass $A_{freeze-out}^{spectator}$, nuclear charge $Z_{freeze-out}^{spectator}$ and excitation energy of the spectator residue after the break-up stage.

In order to calculate mass and atomic number of light clusters emitted in the break-up process, the following considerations are taken:

Many experimental observations established that the production cross-sections in the domain of multifragmentation follow a power law:

$$\frac{d\sigma}{dA} \propto A^{-\tau}, \qquad (48)$$

whose slope is rather well described by an exponent $\tau \approx 2$. The value of $\tau \approx 2$ turned to be rather universal, although a more accurate investigation of experimental data [86] showed a certain dependence on $Z_{bound}$, a quantity often associated to the impact parameter and therefore to the total excitation energy. In ABLA07, the mass of nucleons and fragments produced at break-up is sampled from an exponential distribution with slope parameter $\tau(E^*/A)$, providing that the sampled mass is rejected when exceeding the maximum available mass given as $A_{freeze-out}^{left} = A_{init}^{spectator} - A_{freeze-out}^{spectator}$. The value of $\tau$ is calculated assuming a linear dependence on the excitation energy per nucleon in the temperature regime of interest as discussed in Refs. [87,86]. The sampling is performed several times until the entire mass $A_{freeze-out}^{left}$ is consumed. Each time, the charge $Z_{IMF}$ of the fragment is sampled from a Gaussian distribution centred at $Z_{mean}$, where $Z_{mean}$ is determined by imposing that the ratio $A/Z$ is the same of the hot remnant. The width of the distribution is given by the relation [88]:

$$\sigma_Z^2 = \frac{T_{freeze-out}}{C_{symm}}, \tag{49}$$

where $C_{sym}$ is the symmetry term of the nuclear equation of state. $C_{sym}$ is set to depend on $E^*/A$, as reported in Refs. [89,90,87].

Each of the break-up-fragments greater than an α particle will then enter the evaporation cascade.

In Figure 12 we compare the excitation function for the production of $^7$Be in the reaction $^{93}$Nb+$^1$H calculated with BURST [79] + ABLA07 with experimental data (see [91] and references therein). At lowest proton-beam energies, $^7$Be is produced only via sequential decay from the excited nuclei, while at highest energies also the simultaneous break-up process contributes to its production.

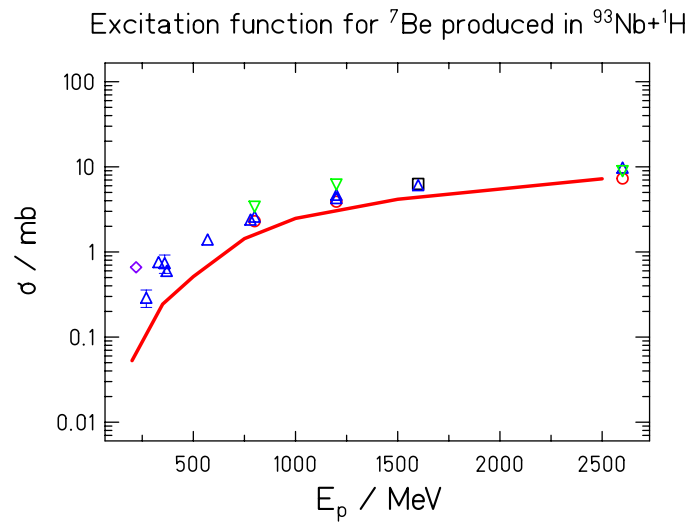

*Figure 12: Excitation function fort the production of $^7$Be in the reaction of $^{93}$Nb+$^1$H – symbols: experimental data (see [92] and references therein), full line: predictions of ABLA07 coupled to BURST [79].*

## III.2. Kinetic-energy spectra

The question on how the fragments acquire their kinetic energies in the multifragmentation process is still vividly discussed, and is closely related to the time scale of the break-up process. If this time scale is very short compared to the time the system needs to reach thermal equilibrium (which at intermediate and high energies is < 100 fm/c [93,94,95]), the break-up system will not reach the thermal equilibrium and dynamical effects play a decisive role, see e.g. [96]. On the contrary, if this time scale is long enough for thermal equilibrium to establish, one can apply statistical considerations as done for example in Ref. [97].

In the first case, the kinematic properties of the created fragments during the break-up are mostly given by the Fermi motion of nucleons in the break-up system. In this case, one can apply the Fermi-gas model [98] for calculating the width $\sigma$ of the momentum distribution of a created fragment:

$$\sigma^2 = \sigma_0^2 \cdot \frac{A_{frag} \cdot \left(A_{init}^{spectator} - A_{frag}\right)}{A_{init}^{spectator} - 1}, \qquad (50)$$

where $A_{frag}$ is the fragment mass, and $\sigma_0$ a parameter amounting to ~118 MeV/c for heavy nuclei. For calculating the kinematical properties, one has to consider two additional effects – influence of thermal motion of nucleons inside the fragment [99] and thermal expansion of the break-up source [24]. Both of these effects will influence the value of the parameter $\sigma_0$ entering Eq. (50).

In the second case, created fragments are in thermal equilibrium with the surrounding gas, and the kinematical properties are mostly given by the thermal motion of fragments inside the break-up volume. In this case, Eq. (50) can be written as [98]:

$$\sigma^2 = m_n \cdot A_{frag} \cdot T_{freeze-out} \cdot \frac{\left(A_{init}^{spectator} - A_{frag}\right)}{A_{init}^{spectator}}, \qquad (51)$$

where $m_n$ is the nucleon mass. Additionally, one has to include the effects of Coulomb repulsion between the nascent fragments in order to calculate their velocities. This is done according to Ref. [100] (see Eq. (4) in Ref. [100]).

Both of these options, i.e. Eqs. (50) and (51) are incorporated in the ABLA07 code, and can be used for calculating kinetic energies of fragments produced in multifragmentation.

## IV. Conclusions

Guided by the empirical knowledge obtained in a recent experimental campaign on the nuclide distributions measured at GSI, Darmstadt, the ABLA code has been subject to important developments. By including the new analytical approximation to the solution of the Fokker-Planck equation for the time dependent fission width, ABLA07 is transformed from a pure statistical code to a dynamical code. It is coupled to the improved semi-empirical fission model PROFI that calculates the characteristics of fragments formed in fission over a large range of energies – from spontaneous fission up to high-energy fission. Apart from neutrons, light charged particles and gammas, also the emission of intermediate-mass fragments is consistently described in ABLA07, thus overcoming the limitation of the previous version of the model in which IMF emission was not considered. The code was originally developed for describing the de-excitation stage of heavy-ion collisions and spallation reactions at relativistic energies. However, coupled to a suitable model for the first stage of the reaction, ABLA07 can also be used to model the de-excitation phase of any kind of nuclear reaction if the approximations of ABLA07 are not considered to be crucial. The parameters of the ABLA07 code are fixed and are the same for all systems and all incident energies, rendering to the code a high predictive power.

## Acknowledgements

We acknowledge the financial support of the European Community under the FP6 Integrated Project EUROTRANS Contract no. FI6W-CT-2004-516520 and "Research Infrastructure Action – Structuring the European Research Area" EURISOL DS Project Contract no. 515768 RIDS. The EC is not liable for any use that may be made of the information contained herein.

# Annex A

In the table below, a comparison between the major physics input of the previous version of the model (ABLA [101,19]) and of the present version (ABLA07) is given.

|  | ABLA | ABLA07 |
|---|---|---|
| Physics Processes | Deexcitation process of a thermalised system – emission of neutrons, protons and $^4$He, and fission | Deexcitation process of a thermalised system – simultaneous break-up, emission of gammas, neutrons, $Z$=1 and 2 particles and intermediate-mass fragments, and fission) |
| Method | Statistical model, Weisskopf formalism | Statistical model, extended Weisskopf formalism |
| Monte Carlo Technique | « timelike » | « timelike » |
| Nuclear level density | Fermi-gas model; Deformation dependence [18]; Energy dependence [18]; Collective enhancement [19] | Fermi-gas model + Constant-temperature model [35,36]; Deformation dependence [18]; Energy dependence [18]; Collective enhancement [19] |
| Coulomb barriers | For protons and $^4$He empirical barriers | For LCP and IMF (all possible species) by nuclear potential [23] plus Coulomb potential; Thermal expansion of the source [$^{102}$] included |
| Nuclear binding energies | Finite-range liquid-drop model including shell and pairing [29] | Finite-range liquid-drop model including shell and pairing [29] |
| Particle-decay width | Geometrical inverse cross sections | Energy-dependent inverse cross sections based on nuclear potential using the ingoing-wave boundary condition model [22]; Tunnelling for LCP included |
| Fission barriers | Finite-range liquid-drop model [64] plus ground-state shell effect [29] | Finite-range liquid-drop model [64] plus ground-state shell effect [29] |
| Angular momentum | Influence of angular momentum on fission barrier is considered | Influence of angular momentum on fission barrier and particle-decay width is considered; Change of angular momentum due to particle evaporation is |

| | | considered |
| --- | --- | --- |
| Dissipation in fission | Transient effect considered by step function | Transient effect considered by approximated solution of the Fokker-Planck equation [25,26]; Influence of initial conditions included [58,59] |
| Low-energy fission probability | Not included | Included according to [1,61] |
| Fission-fragment nuclide distribution | Conditional transition-state model [49,50] | Conditional transition-state model [49,50,51] |

# Annex B

As mentioned above, the correct description of the inverse cross section would lead to the numerical integration of the Eq. (1) and would considerably slow down the calculations.

In fact, using expression (11) the Eq. (1) can be rewritten in the following way:

$$\Gamma_\nu(E_i, J_i) = \frac{2 \cdot s_\nu + 1}{2 \cdot \pi \cdot \rho(E_i, J_i)} \cdot \frac{2 \cdot m_\nu \cdot \pi}{\pi \cdot \hbar^2} \int_0^{E_i - S_\nu - B_\nu} \cdot (R_{geom} + R_\lambda)^2 \cdot \frac{(\varepsilon_\nu - B_\nu)^2}{\varepsilon_\nu} \rho(E_f, J_f) \cdot dE_f. \quad (A1)$$

Following Moretto [103], we can approximate the level density by the constant-temperature formula, with $T$ determined by the inverse logarithmic slope of the level density at the maximum excitation energy of the daughter nucleus. After changing the variable $E_f \rightarrow \varepsilon = \varepsilon_\nu = E_i - E_f - S_\nu$, Eq. (A1) becomes[5]:

$$\Gamma(E_i, J_i) = \frac{2 \cdot s + 1}{2 \cdot \pi \cdot \rho(E_i, J_i)} \cdot \frac{2 \cdot m \cdot \pi}{\pi \cdot \hbar^2} \int_B^{E_i - S_\nu} \cdot (R_{geom} + R_\lambda)^2 \cdot \frac{(\varepsilon - B)^2}{\varepsilon} \rho(E_i - S_\nu - B) \cdot \exp\left(-\frac{\varepsilon - B}{T}\right) \cdot d\varepsilon,$$

$$T = \frac{d \ln \rho(E_f)}{dE_f} \bigg|_{E_f = E_i - S - B}. \quad (A2)$$

This results in three integrals to be solved:

$$I_1 = \int_B^{E_i - S} R_{geom}^2 \cdot \frac{(\varepsilon - B)^2}{\varepsilon} \cdot \exp\left(-\frac{\varepsilon - B}{T}\right) d\varepsilon, \quad (A3)$$

---

[5] For comparison, in the previous version of ABLA the particle decay width was given as:

$$\Gamma(E_i, J_i) = \frac{2 \cdot s + 1}{2 \cdot \pi \cdot \rho(E_i, J_i)} \cdot \frac{2 \cdot m \cdot \pi}{\pi \cdot \hbar^2} \int_B^{E_i - S} \cdot R_{geom}^2 \cdot \varepsilon \cdot \rho(E_i - S - B) \cdot \exp\left(-\frac{\varepsilon - B}{T}\right) \cdot d\varepsilon$$

$$I_2 = \int_B^{E_i-S} 2 \cdot R_{geom} \cdot R_\lambda \cdot \frac{(\varepsilon - B)^2}{\varepsilon} \cdot \exp\left(-\frac{\varepsilon - B}{T}\right) d\varepsilon, \qquad (A4)$$

$$I_3 = \int_B^{E_i-S} R_\lambda^2 \cdot \frac{(\varepsilon - B)^2}{\varepsilon} \cdot \exp\left(-\frac{\varepsilon - B}{T}\right) d\varepsilon, \qquad (A5)$$

*B.1 Inclusion of the Coulomb factor*

The first task is to formulate the decay width for charged particles with the Coulomb factor included in an approximate closed analytical expression (integral $I_1$ in Eq. (A3)). This means that one needs finding the solution of the integral:

$$Y = \int x \frac{x}{x+B} \exp\left(-\frac{x}{T}\right) dx$$

Here, only the general forms are given in order to illustrate the mathematical idea. The variable $x = \varepsilon - B$ is the energy above the barrier, $T$ is the temperature and $B$ is the barrier.

Our basic idea is to calculate the decay width with the combination of two functions:

$$Y = \frac{1}{\frac{1}{Y_1} + \frac{1}{Y_2}}, \quad \text{with} \quad Y_1 = \int x \cdot \exp\left(-\frac{x}{T}\right) dx \quad \text{and} \quad Y_2 = \int \frac{x^2}{B} \cdot \exp\left(-\frac{x}{T}\right) dx.$$

Both integrals can be solved analytically:

$$\int x \cdot e^{ax} dx = \frac{e^{ax}}{a^2}(ax - 1) \quad \text{and} \quad \int x^2 \cdot e^{ax} dx = e^{ax}\left(\frac{x^2}{a} - \frac{2x}{a^2} + \frac{2}{a^3}\right), \quad \text{with} \quad a = -\frac{1}{T}.$$

resulting in:

$$I_1 = R_{geom}^2 \cdot \frac{2T^3}{2T + B}. \qquad (A6)$$

*Justification:*

With $y = x\frac{x}{x+B} \exp\left(-\frac{x}{T}\right)$, $y_1 = x \cdot \exp\left(-\frac{x}{T}\right)$ and $y_2 = \frac{x^2}{B} \cdot \exp\left(-\frac{x}{T}\right)$ we get

$$y = \frac{1}{\frac{1}{y_1} + \frac{1}{y_2}}.$$

Since both curves ($y_1$ and $y_2$) are similar in shape, this relation also holds approximately for the integrals.

## B.2 Inclusion of the energy-dependent quantum-mechanical cross section

In order to include the energy-dependent quantum-mechanical cross section one has to solve the integrals $I_2$ (Eq. (A4)) and $I_3$ (Eq. (A5)).

After replacing $R_\lambda$ in Eq. (A4) with the expression given in Eq. (11), $I_2$ becomes:

$$I_2 = \int_B^{E_i-S} 2 \cdot R_{geom} \cdot k \cdot \frac{1}{\sqrt{\varepsilon}} \cdot \frac{(\varepsilon-B)^2}{\varepsilon} \cdot \exp\left(-\frac{\varepsilon-B}{T}\right) d\varepsilon, \quad k = \sqrt{\frac{\hbar^2 \cdot (M_1+M_2)}{2 \cdot M_2 \cdot (M_1-M_2)}}. \quad (A7)$$

The integral $\int \frac{(\varepsilon-B)^2}{\varepsilon^{3/2}} \cdot \exp\left(-\frac{\varepsilon-B}{T}\right) d\varepsilon$ can be solved analytically [104][6], resulting in:

$$I_2 = 2 \cdot R_{geom} \cdot k \cdot \left[ \frac{2\sqrt{T}(2B^2+TB) + \exp\left(\frac{B}{T}\right)\sqrt{\pi B}(4B^2+4BT-T^2)\text{erf}\left(\sqrt{\frac{B}{T}}\right)}{2\sqrt{TB}} - \frac{\exp\left(\frac{B}{T}\right)\sqrt{\pi}(4B^2+4BT-T^2)}{2\sqrt{T}} \right] \quad (A8)$$

The third integral in Eq. (A5) can be solved analytically in an approximate way like the integral $I_1$:

$$I_3 = \frac{k^2}{\frac{1}{Y_1} + \frac{1}{Y_2} + \frac{1}{Y_3}},$$

where $Y_1 = \int \exp\left(-\frac{x}{T}\right) dx$, $Y_2 = \int \frac{x}{2B} \exp\left(-\frac{x}{T}\right) dx$ and $Y_3 = \int \frac{x^2}{B^2} \exp\left(-\frac{x}{T}\right) dx$

Finally:

$$I_3 = k^2 \cdot \frac{2T^3}{2T^2+4BT+B^2}. \quad (A9)$$

In case of neutrons ($B = 0$), these integrals become:

$$I_1^{neutrons} = R_{geom}^2 \cdot T^2. \quad (A10)$$

---

[6] $\int \frac{(\varepsilon-B)^2}{\varepsilon^{3/2}} \cdot \exp\left(-\frac{\varepsilon-B}{T}\right) d\varepsilon = -\frac{\exp\left(\frac{B-\varepsilon}{T}\right)\left[2\sqrt{T}(2B^2+T\varepsilon) + \exp\left(\frac{\varepsilon}{T}\right)\sqrt{\pi\varepsilon}(4B^2+4BT-T^2)\text{erf}\left(\sqrt{\frac{\varepsilon}{T}}\right)\right]}{2\sqrt{T\varepsilon}}$

$$I_2^{neut} = R_{geom} \cdot k \cdot \sqrt{\pi} T^{3/2}. \quad (A11)$$

$$I_3^{neut} = k^2 \cdot T. \quad (A12)$$

In Figure A.2 we show the ratio between the analytical approximation of Eq. (1), given by Eqs. (A6 – A9) for light charged particles and Eqs. (A10 – A12) for neutrons, and the result of the numerical integration of Eq. (1). This ratio is shown for several different nuclei. We see that in case of neutron there is no difference between the analytical and the numerical solutions. In case of light charged particles, the analytical approximation over-estimates the particle width by less than 10 % as compared to the numerical solution. Thus, we can conclude that the analytical approximation to the Eq. (1) is quit realistic in calculating the particle-decay width.

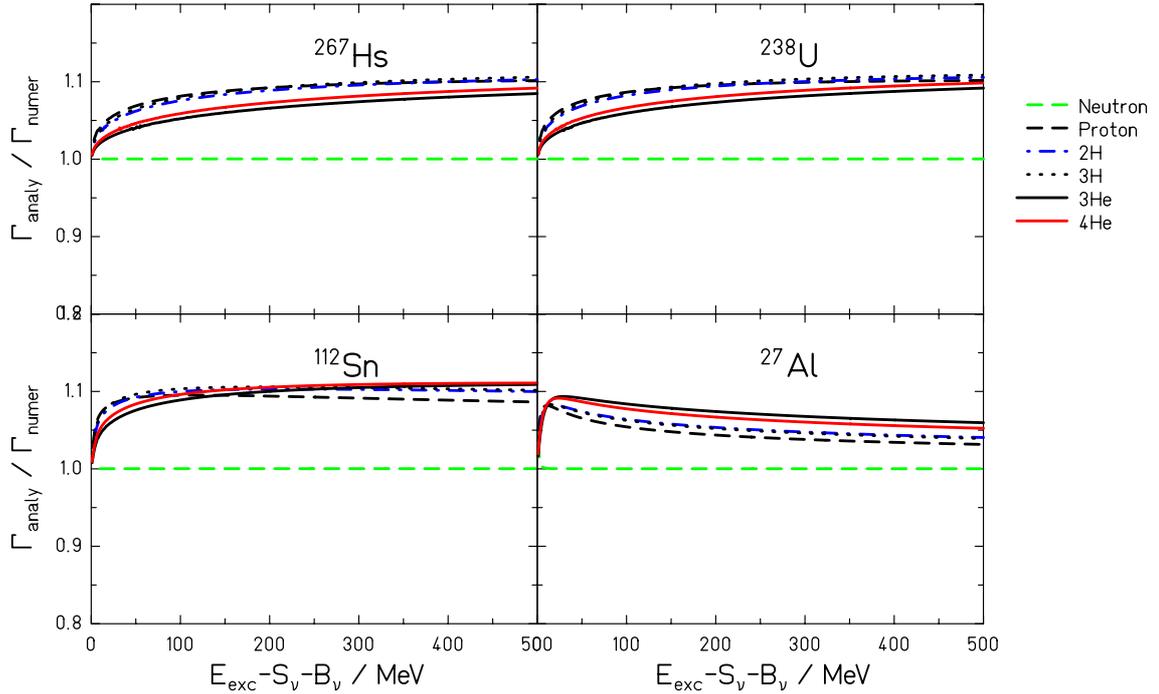

*Figure A.1: Ratio between numerical and analytical solution of Eq. (A.1) as a function of excitation energy for the case of neutron and light-charged particle emission from different nuclei.*